\documentstyle[psfig,sprocl]{article}

\def\be{\begin{equation}}
\def\ee{\end{equation}}
\def\bea{\begin{eqnarray}}
\def\eea{\end{eqnarray}}

\begin{document}

\title{ $J/\Psi $ ABSORPTION SCENARIOS IN NUCLEAR COLLISIONS }

\author{J. GEISS \footnote{Talk presented at the
INT/RHIC workshop `Charmonium Production in Relativistic Nuclear Collisions'
Seattle (USA), May 1998}, E. BRATKOVSKAYA, W. CASSING and  C. GREINER  }

\address{Institut f\"ur Theoretische Physik, Universit\"at Giessen,\\
D-35392 Giessen, Germany }

\maketitle\abstracts{
We study the production of $c \bar{c}$ pairs and dimuons from hard
collisions in nuclear reactions within the covariant transport approach
HSD. Adopting 6~mb for the $c
\bar{c}$-baryon cross section the data on $J/\Psi$ suppression in
p~+~A reactions are reproduced in line with calculations based on the
Glauber model. Furthermore, using $J/\Psi$ absorption cross sections with
mesons above the $D\bar{D}$ threshold in the order of 1.5 - 3~mb we
find that all data on $J/\Psi$ suppression from NA38/NA50
can be described without assuming the formation of a quark-gluon plasma.
Alternatively, we also
investigate an 'early'-comover absorption scenario where the
$c \bar{c}$ pairs are dissociated in the color electric fields
of neighboring strings.
Again we find good agreement with the experimental
data with an estimate for the string
radius of $R_s \approx 0.2-0.25 \,fm$. 
}

\section{Introduction}

More than a decade ago
Matsui and Satz \cite{matsui} have proposed that a suppression of the
$J/\Psi$ yield in ultra-relativistic heavy-ion collisions is a
plausible signature for the formation of the quark-gluon plasma because
the $J/\Psi$ should dissolve in the QGP due to color screening.
This suggestion has stimulated a number of
heavy-ion experiments at CERN SPS to measure the $J/\Psi$ via its
dimuon decay. Indeed, these experiments have shown a significant
reduction of the $J/\Psi$ yield when going from proton-nucleus to
nucleus-nucleus collisions \cite{NA38}. Especially for Pb~+~Pb at
160~GeV/A an even more dramatic reduction of $J/\Psi$ has been reported
by the NA50 collaboration\cite{NA50,gonin}.

To interpret the experimental results, various models based on $J/\Psi$
absorption by hadrons have also been proposed \cite{gonin,gerschel,KhaQM96}.
However, these models have failed to
explain the ``anomalous" suppression reported in central Pb~+~Pb
collisions, thus leading to the suggestion of a possible formation of a
quark-gluon plasma in these collisions \cite{gonin,KhaQM96,blaizot,wong}.
On the other hand, Gavin and Vogt \cite{gavin,gavin1}, based also
on the hadronic absorption model, have found that although $J/\Psi$
absorption by nucleons is sufficient to account for the measured total
$J/\Psi$ cross sections in both proton-nucleus and nucleus-nucleus
collisions, it cannot explain the transverse energy dependence of
$J/\Psi$ suppression in nucleus-nucleus collisions.  To account for the
nucleus-nucleus data they have introduced additionally the absorption
on mesons (`comovers') with a cross section of about 3~mb. A similar
model has also been proposed by Capella et al. \cite{Capella} to describe
the $J/\Psi$ and $\Psi^\prime$ suppression in nucleus-nucleus collisions.

In all these studies the dynamics of the collisions is based on the
Glauber model, thus a detailed space and time evolution of the colliding
system is not included. In the following we want to report
on our theoretical analysis of $J/\Psi$ production and absorption
within a transport theoretical analysis.
On the one hand we will show our detailed microscopic analyis of the
(hadronic) comover scenario. As an alternative we also consider
the effect of $c\bar{c}$ dissociation in
the prehadronic phase of the heavy-ion collision, which can be regarded
as an 'early'-comover absorption scenario,
and not as absorption by 'late' hadronic comovers.
This is motivated by the fact, that the very early collision phase is 
not described by hadrons but by highly excited strings.
As each individual string carries a lot of internal energy (to produce the later secondaries)
in a small and localized space-time volume
the quarkonia state might get completely dissociated by the
intense color electric field inside a single string \cite{loh}.

\section{Ingredients of the transport calculations}

Within our transport calculation we can exploit
various assumptions (models) for the $c\bar{c}$ formation and propagation
and also take into account the Drell-Yan process
explicitly. This represents a sophisticated extension of the first analysis
carried out in Ref.~\cite{Ca97}.
A detailed description of the transport approach as well as
its ingredients is given in Refs.~\cite{Ca97b,Cassing}.
The generation of Drell-Yan events is performed
with the PYTHIA event generator \cite{PYTHIA} version 5.7 using GRV LO
or MRS mode 43 structure functions from the PDFLIB package \cite{PDFLIB}.

According to our dynamical prescription the Drell-Yan pairs can be
created in each hard $pp, pn$ or $nn$ collision ($\sqrt{s} \geq$
10~GeV).  Since PYTHIA calculates the Drell-Yan process in leading
order only (using GRV LO structure functions) we have multiplied the
Drell-Yan yield for $NN$ collisions by a K-factor of 2.0
(cf. Refs.~\cite{NA38,Carlos}).  In case of MRS mode 43 structure functions, which
include NLO corrections, a K-factor of about 1.6 had to be
introduced (cf. Ref.~\cite{CarQM96}).
We have compared our results within the production scheme given above
with that used in Refs.~\cite{NA38,NA50,Carlos}, where the Drell-Yan
yield from p~+~A and A~+~A collisions is calculated as the isotopical
combination of the yield from $pp$ and $pn$ at fixed $\sqrt{s_0}$
scaled by $A_P \times A_T$. We found that the variation from the scheme
used in Refs.~\cite{NA38,NA50,Carlos} is less then 10\%.
As the production scheme is considered to be the same for $c\bar{c}$-pairs their
total cross section (without reabsorption) also scales with $A_P \times
A_T$; the ratio of the $J/\Psi$ to the Drell-Yan cross section thus
provides a direct measure for the $J/\Psi$ suppression.
We emphasize
that the $A_P \times A_T$ scaling of the initial $c\bar{c}$ production
is an input of our calculation and not a result.

We then follow the motion of the $c \bar{c}$ pair in hadronic matter
throughout the collision dynamics by propagating it as a free particle.
In our simulations the $c\bar{c}$ pair, furthermore, may be destroyed
in collisions with hadrons using the minimum distance concept as
described in Sec. 2.3 of Ref.~\cite{Wolf90}.  For the actual cross
sections employed we study two models (denoted by I and II) which both
assume that the $c\bar{c}$ pair initially is produced in a color-octet
state and immediately picks up a soft gluon to form a color neutral
$c\bar{c}-g$ Fock state\cite{KhaQM96} (color dipole).  This extended
configuration in space is assumed to have a 6~mb dissociation cross
section in collisions with baryons ($c \bar{c}+B\to\Lambda_c+\bar D$)
as in Refs.~\cite{gonin,gerschel,KhaQM96} during the lifetime $\tau$
of the $c\bar{c}-g$ state which is a parameter. In the model I we
assume $\tau = $ 10~fm/c, which is large compared to the nucleus-nucleus
reaction time such that the final resonance states $J/\Psi$ and
$\Psi^\prime$ are formed in the vacuum without further interactions
with hadrons. In the model II we adopt $\tau =$ 0.3 fm/c as suggested
by Kharzeev \cite{KhaQM96} which implies also to specify the
dissociation cross sections of the formed resonances $J/\Psi$ and
$\Psi^\prime$ on baryons. For simplicity we use 3~mb following
Ref.~\cite{KhSatz96}.
Within the comover scenario the cross section for $c\bar{c}-g, J/\Psi$, or
$\Psi^\prime$ dissociation on mesons ($c\bar{c}+m\to D\bar{D}$) is
treated as a free parameter ranging from 0 to 3~mb.

\begin{figure}[ht]
\centerline{\psfig{figure=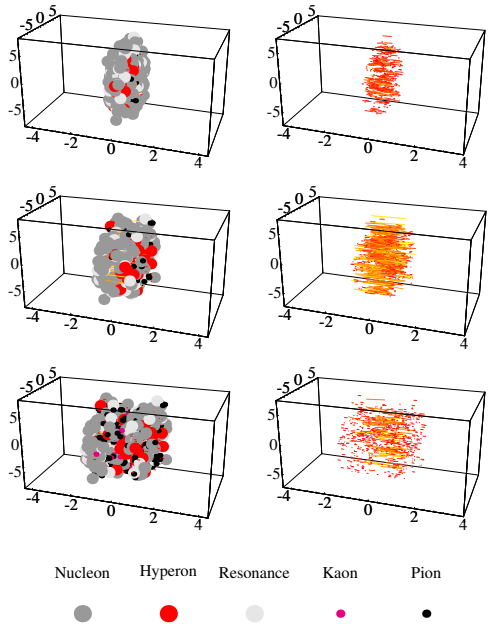,height=10cm}}
\caption{
Graphical representation of the hadrons (left) 
and the strings (right)
during the high density phase in a central 
Pb~+~Pb collision at 160 AGeV in the center-of-mass system.
Three time steps are shown at $t_{cm}=(t_0, t_0+0.6, t_0+1.2)$ fm/c;
the axis labels are given in fm.
}
\end{figure}

In our alternative approach we discard the possible absorption on late
mesonic comovers, but instead focus on the possible
$c\bar{c}$ dissociation by the strings in
the prehadronic phase \cite{geiss}.
The absorption on the surrounding nucleons will be treated
like for model II.
In the early stages of the reaction the $c\bar{c}$ states
move not in a hadronic environment but in an environment of color electric
strings of 'wounded' nucleons:
Several hundred strings are formed during a central Pb-Pb collision at SPS energies.
In Fig.~1 a characteristic representation of the hadrons
and the strings during the high density phase in a central 
Pb~+~Pb collision at 160 AGeV is shown. 
The high energy hadron-hadron 
collisions are described by the FRITIOF model \cite{LUND} 
resulting in two excited 
strings.
The dynamical evolution of the strings is now
included explicitly (for details see Ref.~\cite{geiss}).
The fragmentation of the strings into hadrons starts after the 
formation time, which is set to  $t_f=0.8$ fm/c.
We now
assume that a $c\bar{c}$ state immediately
dissociates whenever it moves into the region of the color electric 
field of a string. In this sense strings are completely 
'black' for $c\bar{c}$ states. It was found in Ref.~\cite{loh} that the additional force
acting on the charm quarks in the color electric field
is given by $2 \times \sigma \approx 2 \,GeV/fm$,
where $\sigma$ denotes the phenomenological string constant of a 
chromo electric flux, which is sufficient to immediately break up a
$c\bar{c}$ state. One can also argue that the field energy density contained in a 
string is given by $\sigma / (\pi R_S^2)$. For $R_S \approx 0.3 \,fm$ one
accordingly has a local high color electric energy density of 
$\approx 4\, GeV/fm^3$, which substantially screens the binding potential
of the charmonium state \cite{matsui}. 
For practical reasons the dissociation by a string is modeled
when the center-of-mass of the $c\bar{c}$ state is located inside the string
of radius $R_s$.
(Absorption by strings spanned between the parent particles of the
$c\bar{c}$ pair is excluded, since this effect already 
is included in the production cross section.)
The string radius $R_s$ represents an unknown parameter.

\section{Comparison to experimental data}

Since the absorption of $c\bar{c}$-pairs on secondary mesons in
proton-nucleus collisions is practically negligible \cite{Ca97}, these
reactions allow to fix `experimentally' the $c\bar{c}$-baryon
dissociation cross section on nucleons. Our analysis in Ref.~\cite{Ca97b}
yields $\sigma^{abs}_{baryons}\, \approx $ 6 mb. The experimental survival
probability is defined by the ratio of experimental $J/\Psi$ to Drell-Yan cross
section as
$$S_{exp} = \left.{\left(B_{\mu\mu}\sigma^{J/\Psi}_{AB}\over
\sigma^{DY}_{AB}|_{2.9-4.5 \ {\rm GeV}}\right)} \right/
{\left(B_{\mu\mu}\sigma^{J/\Psi}_{pd}\over \sigma^{DY}_{pd}\right)},$$
where $A$ and $B$ denote the target and projectile mass while
$\sigma^{J/\Psi}_{AB}$ and $\sigma^{DY}_{AB}$ stand for the $J/\Psi$ and
Drell-Yan cross sections from $AB$ collisions, respectively, and
$B_{\mu\mu}$ is the branching ratio of $J/\Psi$ to dimuons.  We note
that due to the large statistical error bars of the experimental data
absorption cross sections $\sigma_{abs}^{baryons}$ of 6 $\pm$ 1~mb
are compatible also. These values are slightly smaller than those of
Kharzeev et al.~\cite{KLNS96} in the Glauber model claiming
7.3~$\pm$~0.6~mb, but in the same range as those used in the Glauber
models of Ref.~\cite{gerschel}.

The comparison to $J/\Psi$ suppression in nucleus-nucleus collisions
is performed on an event-by-event basis using the neutral transverse
energy $E_T$ (cf. Ref.~\cite{Ca97b}) as a trigger as in the experiments of
the NA38 and NA50 collaborations.  We first compute the results for S +
U at 200~GeV/A and Pb~+~Pb at 160~GeV/A within the model I varying the
dissociation cross section on mesons of the $c\bar{c}-g$ object from 0
to 1.5~mb while keeping the absorption cross section on baryons fixed
at 6~mb.  The calculated $J/\Psi$ survival probabilities are displayed in
Fig.~2 (l.h.s.) in comparison to the data from Ref.~\cite{NA50}
for both systems; the dashed
lines are obtained for $\sigma^{mesons}_{abs}$ = 0 mb while the solid
lines correspond to $\sigma^{mesons}_{abs}$ = 1.5 mb.  Whereas the data
for S~+~U appear to be approximately compatible with our calculations
without any dissociation by mesons, the Pb~+~Pb
system shows an additional suppression. This finding is in agreement
with the results of Glauber models \cite{KhaQM96,KLNS96}.  On the other
hand, the Pb~+~Pb data are well reproduced with a cross section of
1.5~mb (in model I) for the $J/\Psi$ absorption on mesons which,
however, then slightly overestimates the suppression for the S~+~U data
for the 3 middle $E_T$ bins.

\begin{figure}[ht]
\centerline{\psfig{figure=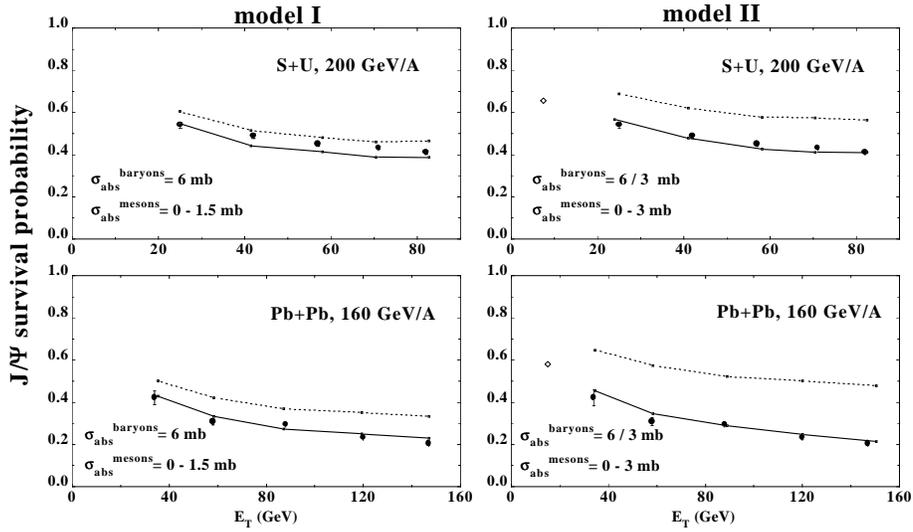,height=7cm}}
\caption{
The $J/\Psi$ survival probability for S~+~U at 200~GeV/A (upper
part) and Pb~+~Pb at 160~GeV/A (lower part) as a function of the
transverse energy in comparison to the experimental data from
Ref.~\protect\cite{NA50} within the model I assuming a long lifetime for
the $c\bar{c}-g$ system (l.h.s.) as well as within model II (r.h.s.) for
a respective lifetime of 0.3 fm/c.  The absorption cross section on mesons is
varied from 0 (dashed lines) to 1.5~mb or 3.0 mb, respectively (solid lines).
}
\end{figure}

We, furthermore, compute the $J/\Psi$ suppression for S~+~U at
200~GeV/A and Pb~+~Pb at 160~GeV/A within the model II for a
$c\bar{c}-g$ lifetime $\tau$ = 0.3~fm/c varying the dissociation cross
section of the $c\bar{c}-$pair with mesons from 0 - 3~mb while keeping
the absorption cross section on baryons fixed at $\sigma_{abs}^{baryons}$ =
6~mb for the `pre-resonance' state and at 3~mb for the formed $J/\Psi$
resonance.  The calculated survival probabilities are displayed in
Fig.~2 (r.h.s.) in comparison to the data; again the dashed lines
correspond to the calculations without any charmonium absorption on
mesons whereas the solid lines represent our calculations for a meson
absorption cross section of 3~mb. In the absorption model II the data
for S~+~U are no longer compatible with our calculations without any
dissociation by mesons.  The S~+~U data here need an absorption by
mesons in the range of 3~mb as in the model of Gavin
et al.~\cite{gavin,gavin1}.

We now describe our results within the `early'-comover scenario.
For p~+~U and $R_S=0.4$ fm only 2\% of the $J/\Psi$'s  
are absorbed by strings. The absorption is thus dominated, as expected
intuitively, by the $c\bar{c}$-baryon
dissociation on nucleons.
This turns out to be completely different for heavy-ion collisions,
where the absorption on strings becomes a much more important effect.

\begin{figure}[ht]
\centerline{\psfig{figure=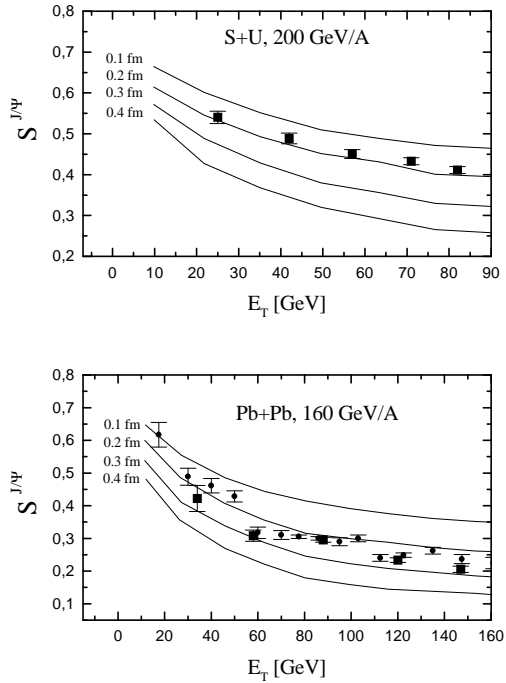,height=9cm}}
\caption{The $J/\Psi$ survival probability $S^{J/\Psi}$ for S~+~U at 200~
A$\cdot$GeV (upper
part) and Pb~+~Pb at 160~A$\cdot$GeV (lower part) as a function of the
transverse energy $E_T$ in comparison to the experimental data from
Ref.~\protect\cite{NA50} (full squares).
The full circles in the lower part show the new NA50 data from
Ref.~\protect\cite{NA50a}.
The calculated results are shown for the 
string radii $R_s=$0.1, 0.2, 0.3 and 0.4 fm.
}
\end{figure}

In Fig.~3 our results are shown for  S~+~U and Pb~+~Pb as a function of
the transverse energy and for different string radii $R_s$ = 0.1,...,0.4 fm. 
A strong dependence
on the string radius $R_s$ is observed with $R_s \approx$ 0.2 fm giving 
the best fit 
to the experimental data \cite{NA50,NA50a}. For this string radius 40\% of the absorbed 
$J/\Psi$'s are dissociated by strings in central collisions 
of Pb~+~Pb.

In principle the mesonic comover absorption
scenario could be included in addition;
however, since most of the $c\bar{c}$ pairs are already absorbed by
strings and baryons in the early stage of the reaction, the effect
of meson comovers is expected to be much less compared
to the pure hadronic comover scenario (model II) discussed above.

\section{Summary}

In this contribution we have summarized the results of
microscopic transport studies of
$J/\Psi$ and Drell-Yan production in proton-nucleus and
nucleus-nucleus collisions \cite{Ca97b,geiss} within different
scenarios.

Our first set of calculations show that the absorption
of `pre-resonance' $c\bar{c}-g$ states by both nucleons and produced
mesons (the standard, i.e. late hadronic `comover'-scenario)
can explain reasonably not only
the inclusive $J/\Psi$ cross
sections but also the transverse energy ($E_T$) dependence of $J/\Psi$
suppression measured in  nucleus-nucleus collisions.  In particular,
the absorption of $J/\Psi$'s by produced mesons is found to be
important especially for Pb~+~Pb reactions,
where the $J/\Psi$-hadron reactions extend to longer times as
compared to the S~+~W or S~+~U reactions.  This is in contrast with results
based on a simple Glauber model, which neglects both the transverse
expansion of the hadronic system and the finite meson formation times.

In our second set of calculations
we found that the early absorption by strings (the `early'-comover scenario)
is an important and dominant effect in the
first few fm/c of the collision phase before secondary particles
are produced. Adopting a string radius of 0.2-0.25 fm we got a
similar good quantitative
agreement with p+A data and the NA38 and NA50 data
by taking into account the absorption by strings as well as nucleons.
This radius seems to
be rather small, but for two reasons it should be seen as a lower bound 
for $R_s$.
First of all, we have assumed that the strings are completely 'black' for  
$c\bar{c}$ states, and secondly, the $c\bar{c}$ pair  
dissociates whenever it moves into the region of the color electric 
field of a string. With the requirement
that the center of the $c\bar{c}$ state should be inside the string for the 
dissociation
process to start, one should add the $c\bar{c}$ radius to our value of $R_s$;
this then gives a string radius of $\tilde{R}_s\approx 0.4-0.5$ fm.

As a result of our transport studies we do not find a necessary
argument to require the formation of a quark-gluon-plasma in Pb~+~Pb
collisions at the SPS.

\section*{Acknowledgments}
This work was supported by BMBF and GSI Darmstadt.

\end{document}